\documentclass[11pt]{article}
\usepackage{hyperref}
\pdfoutput=1

\begin{document}
\title{DPIV Measurements of Olympic Skeleton Athletes (Fluid Dynamics Video)}
\author{Chia Min Leong$^1$, YaeEun Moon$^1$, Vicki Wu$^1$, Timothy Wei$^1$ and\\ Steve Peters$^2$ \\
\\\vspace{6pt} $^1$ Rensselaer Polytechnic Institute, Troy, NY 12180, USA
\\\vspace{6pt} $^2$ USA Bobsled \& Skeleton Federation }
\maketitle

The Olympic sport of skeleton involves an athlete riding a small sled face first down a bobsled track at speeds up to 130 km/hr. In these races, the difference between gold and missing the medal stand altogether can be hundredths of a second per run. As such, reducing aerodynamic drag through proper body positioning is of first order importance. To better study the flow behavior and to improve the performance of the athletes, we constructed a mock section of a bobsled track which was positioned at the exit of an open loop wind tunnel. DPIV measurements were made along with video recordings of body position to aid the athletes in determining their optimal aerodynamic body position. In the fluid dynamics video shown, the athlete slowly raised his head while DPIV measurements were made behind the helmet in the separated flow region.  
%
\end{document}